\documentclass[doublecol]{epl2}
\title {Quantum Phase Transitions in a Dimerized Bose-Hubbard Model: A DMRG Study}

\author{ Anasuya Kundu and Swapan K Pati$^\dag$}

\institute{Theoretical Sciences Unit and DST Unit on Nanoscience\\
Jawaharlal Nehru Center for Advanced Scientific Research\\
Jakkur Campus, Bangalore 560064, India\\
$^\dag$E-mail: pati@jncasr.ac.in}


\abstract{
We investigate the phase
diagram of a dimerized Bose-Hubbard model, using density matrix renormalization group
technique. We find a new phase, which is the coexistence of
superfluid and bond-wave phases, due to the effect of dimerization. 
Experimentally dimerization in optical lattice can be realized by using two
counter propagating laser beams of different wavelengths.
Apart from the conventional superfluid to Mott insulator transition,
we find a new quantum phase transition:
from superfluid-bond-wave to Mott insulator-bond wave phase.
Our study suggests a rich phase diagram which can be easily probed. }

\begin{document}

\maketitle

\section{Introduction}
Over the last decade, there have been a large number of interesting theoretical predictions on
the rich physics of strongly correlated optical lattice
systems~\cite{Jaksch, Batrouni, Kovrizhin}.
The spectacular experiment by Greiner {\it et al.}~\cite{Greiner} has opened a new frontier,
as it convincingly demonstrates that at low temperature, the cold atoms trapped in an
optical lattice undergo a quantum phase transition from the delocalized superfluid (SF)
state to the localized Mott insulator (MI) state. The unprecedented control in experiment
over the theoretical parameters provides a good opportunity to study the various
competing phases in this class of lattice systems.

The theory of the underlying quantum phase transition of strongly interacting bosons
confined in an optical lattice is described by the
well-known Bose-Hubbard (BH) model~\cite{Fisher}. The BH Hamiltonian
is given by
\begin{equation}
H_{BH}=-t\sum_{<ij>}\left(b_i^\dag b_j + h.c.\right) + U\sum_i n_i(n_i-1)/2
\end{equation}
where $t$ is the hopping parameter, $b_i$ ($b_i^\dag$) are the annihilation (creation) operators of
bosons on site $i$ and $n_i=b_i^\dag b_i$ represents the number of bosons on site $i$.
$U$ is the repulsion between a pair of bosons occupying the same site.
Experimentally, the ratio $t/U$ is
varied by tuning the laser intensity, thus the depth of the optical potential
can be adjusted. When $U/t << 1$ {\it i.e.}, the optical trap is very shallow,
the hopping between different lattice sites dominates, making the atomic wave functions 
completely delocalized, giving rise to a superfluid phase.
As $U/t$ is increased, the depth of the
optical trap increases and the system is driven towards Mott insulator phase.
In this regime, since fluctuations in the boson number become energetically costly,
bosons localize at different lattice sites. Thus there
are fixed number of bosons at each site. 

The BH Hamiltonian has been studied extensively using mean-field
theory~\cite{Fisher, Sheshadri, Kovrizhin}, quantum Monte Carlo
techniques~\cite{Batrouni, Krauth, Niyaz, Kashurnikov}
and density matrix renormalization group (DMRG)~\cite{White, Schollwock}
method. DMRG is a very
promising and powerful numerical technique for low dimensional interacting systems.
This gives accurate results with a very good precision at low temperature.
It is useful for finding accurate approximations of the ground state
and the low-lying excited states of strongly interacting quantum lattice systems.
DMRG has been proved to be successful for
bosonic~\cite{Rapsch, Kuhner, Kuhner1}, 
fermionic~\cite{Whitefm} and spin~\cite{Huse, Noack} systems.

\section{Dimerized BH model}
We employ DMRG method to study one-dimensional (1D) system of interacting bosons in 
presence of onsite repulsion and hopping dimerization. The basic physics which governs the 
bosons trapped in optical lattice is the competition between the kinetic energy and the 
onsite repulsive potential energy. The kinetic energy can be controlled by laser field
modulation. If two counter propagating laser beams of different wavelengths are used,
the hopping amplitude can be modulated as weaker and stronger in
consecutive sites in a lattice. This is because the laser induced optical trap can
be tuned with the wavelengths of the counter propagating laser beams so that they generate
deeper and shallower traps in alternate optical lattice sites. For such a dimerization
in hopping strengths, while the Hubbard $U$ term localizes the particles in lattice sites, 
dimerization would lead to an insulating phase. 
In fact, the dimerization is one reason why no one-dimensional lattice system
can be metallic~\cite{Peierls}.
In one-dimension, for Fermionic systems, the potential energy
$\frac{1}{2}kx^2$ from the
springs competes with the dimerization.
Whether tuning of well depth stabilizes a one-dimensional optical lattice and
what would be the
effect of dimerization on superfluid to Mott insulator transition have not been studied before.

In this letter, we consider dimerization in hopping integral to describe basic physics
associated with such a term. The Bose-Hubbard model with dimerization can be written
as   
\begin{equation}
H=\sum_i\left[t+(-1)^{i+1}\delta\right]\left(b_i^\dag b_{i+1} + h.c.\right) +
\frac{U}{2}\sum_i n_i(n_i-1)
\end{equation}
where $\delta$ is the dimerization parameter which creates hopping strengths $t+\delta$ and $t-\delta$
at alternate sites keeping the total length of the one-dimensional
lattice the same. 
We set $t=1$ in all our calculations and express energy in units of $t$.

In DMRG, with open boundary condition, we vary the density matrix cut-off so as to get
the converged result. Although, there is no restriction on the maximum number
($n$) of bosons per site, for computational purpose $n$ has to be
truncated at a finite value. We choose $n=3$, {\it i.e.}, with $0$, $1$ and $2$
bosons per site. We have checked the convergence of our results with $n=4$ and
$5$ also. In literature, estimation of critical point ($U_c$) for SF to MI transition
has been obtained by using different methods.
K\"uhner {\it et. al.}~\cite{Kuhner1}
have used finite-size DMRG with open boundary condition and estimated $U_c$ as $3.37 \pm 0.1$
by studying the decay of the correlation function.
There are also other methods like Bethe Ansatz~\cite{Krauth1}
($U_c \simeq 3.46$), exact diagonalization and renormalization group~\cite{Kashurnikov}
($U_c \simeq 3.29 \pm 0.01$) and Quantum Monte Carlo~\cite{Kashurnikov1}  
($U_c \simeq 3.33 \pm 0.05$) by which the critical point has been estimated.

In this letter, we focus on the $\delta$ vs. $U$ phase
diagram. The interesting physics which 
is revealed due to the effect of hopping dimerization, is the appearance of
a new phase;
bond-wave (BW) phase, coupled with the existence of conventional
superfluid and Mott insulator phases.
There are many suggestions and examples of supersolid (SS) phases where superfluid
order coexists with modulation
of the boson density~\cite{Kovrizhin, Batrouni, Kim}. Our central result points towards
existence of a new phase where we find simultaneous presence of superfluid order and 
modulation of bond kinetic energy.
We can draw an analogy of our model with the Boson-Hubbard model with
an additional `superlattice' potential, in which the confining potential has multiple
minima~\cite{Rousseau}. In some
recent experiments, the `superlattice' potential
has already been implemented~\cite{Bloch, Bloch1, Bloch2}.  
Similar to the consequence of dimerization which results in a bond wave order that breaks
translation invariance, the superlattice potential likewise causes the formation of a
density order pattern which can compete with or coexist with a
superfluid phase~\cite{Rousseau}.

In the Fermionic Hubbard model
at half-filling, there are examples
of coexistence of density wave (DW) with bond-wave
phases~\cite{Ung}. In this context, since
kinetic energy controls density alternation, its modulation leads to BW
phase  and in a periodic system,
both DW and BW order parameters are equivalent in the momentum space.
We find that along the $U$-line 
($\delta=0$), the phase transition is from superfluid to Mott insulator phase,
while along the $\delta$=1 line the superfluid phase vanishes,
since the dimerized hopping breaks the optical lattice into disconnected dimers.
We also estimate the critical line and the complete phase diagram in the two-dimensional 
$U-\delta$ parameter space.

\section{Results and Discussions}
To obtain a clear insight into the complete phase
diagram in the $U-\delta$ plane, we have calculated
ground state energy, excitation gap and various correlation functions. For the well known BH Hamiltonian 
with $\delta=0.0$, we find the excitation to be gapless in the superfluid phase, whereas a small gap opens 
up at the SF to MI transition point as found previously. The critical point, however,
can not be solely determined from the gap and since inclusion of $\delta$ opens up a dimerized gap,
we define an order parameter for the SF phase, the asymptotic value of the quantity $<b_i^\dag b_j>^{1/2}$
as $|i-j| \rightarrow \infty$, and estimate the critical $U_c$ at
which this quantity vanishes. For $\delta=0$, 
the critical $U$-value is found to be $3.37 \pm 0.1$ in agreement with~\cite{Kuhner1}.

To bring out the effect of dimerization, we first study bond-wave phase. The bond-wave phase is
characterized by the alternating strengths of the expectation value of the
kinetic energy operator on the bonds. This quantity, $<b^\dag_ib_{i+1} + h.c.>$, against the bond 
index, $i$, is shown in Fig.~1(a) for a range of $U$ values and for $\delta=0.1$ and $\delta=0$. 
As can be seen, for $\delta=0$,
there is equal strength of kinetic energy on each bond, but for finite value of $\delta$, the 
kinetic energy strength alternates characterizing the BW phase. Interestingly, the alternation is
found to be stronger for larger $U$ values. In Fig.~1(b), we plot the thermodynamic stabilization 
energies per site
with respect to the undimerized system against $\delta$ for small and large U values.
For all the cases, we find that the stabilization energy varies as $A\delta^2$, where $A$ depends on $U$.
The prefactor $A$ turns out to be $0.158$, $0.252$ and $0.267$ for $U=0.5$, $3.0$ and $5.0$ respectively.
This shows that the stabilization energy gained by the system is of the same power
in distortion as the lattice term. Thus, the stabilization is conditional and depends crucially 
on the width of the optical trap potential. Moreover, for a particular value of $\delta$,
the stabilization energy is more for larger $U$ values.
The SF phase is characterized by fluctuations in particle numbers, while in MI phase,
in a given site, the particle number is fixed as it becomes energetically costly for the particle to
hop. To understand such a variation, we plot the local density, $<n_i>$, as a function of lattice
site in the inset (of Fig.~1(b)) for small and large $U$ values and
for two different $\delta$s.
As can be seen, when the system is in SF phase, it is easier for the bosons to
hop between sites, hence there is a large fluctuation of this quantity. On contrary, in MI phase
the bosons are localized and so the fluctuation in $<b^\dag_ib_i>$ reduces with increase in correlation
strength. It is also interesting to note that with increase in dimerization strength, the
particle number fluctuations reduce considerably, giving an indication that even $\delta$ drives the
system out of SF phase.

\begin{figure}
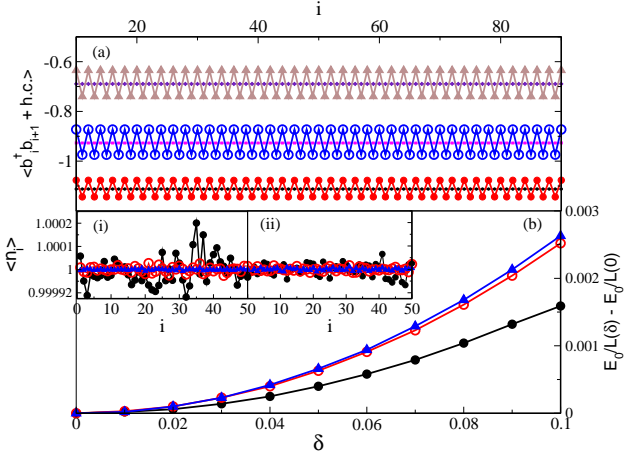

\onefigure[scale=0.3]{fig1.eps}
\caption{(color online) (a) Plot of expectation values of kinetic energy operator against
the bond index for $\delta=0.1$ and $U=0.5$ (filled circle), 
$3.0$ (open circle)
and $5.0$ (triangle). The points along the middle of each present the same for
$\delta=0$.
(b) Variation of stabilization energy against $\delta$ for $U=0.5$ (filled circle),
$3.0$ (open circle) and $5.0$ (triangle). The points are joined by lines to guide the eye.
Inset: Plot of local density against the site index ($i$) for (i) $\delta=0.0$ and
(ii) $\delta=0.5$. In each case the $U$ values are $0.5$ (filled circle),
$3.0$ (open circle) and $5.0$ (triangle). } 
\end{figure}

The other quantity that varies considerably near the SF to MI phase transition is the
extent to which the lattice sites are correlated. Whether dimerization also reduces the 
correlation length, we study the hopping correlation function, $<b_0^\dag b_r>$,
with the distance $r=|i-j|$ for a range of values in the $U-\delta$ plane. For three $U$ values,
we plot $<b_0^\dag b_r>$ for the undistorted as well for system with increasing dimerization
in Fig.~2. As can be seen, when the system is in superfluid phase, {\it i.e.}, for small $U$, the 
correlation function shows a quasi long range order. In this regime, the decay of the correlation
function obeys a power law behavior $\sim r^{-K/2}$ ($K$ is the Luttinger liquid parameter).
In SF phase we find $K \ge 2$.
As it is clear from the figure, with increase in onsite repulsion, {\it i.e.},
when the system transits to the MI phase, the decay of the correlation function
becomes faster.
In MI phase, the particles are pinned at each lattice sites, so the ordering becomes short
ranged. We find that in MI phase, the decay of the correlation function
is exponential, which is given by $e^{-r/\xi}$; $\xi$ being the correlation length.
For $U=5.0$, this $\xi$ comes out to be 
2.38, 2.35, 2.34, 2.10, 1.68 and 1.46 for $\delta = 0.0$, 0.05, 0.1, 0.3, 0.5
and 0.7 respectively. So there is a gradual decrease of correlation length
with increase in dimerization. It is to be noted that
with increase in dimerization, the decay of the correlation function becomes faster 
and in Fig.~2(c) even for $U=0.5$, the decay of the correlation function
is exponential. Thus, as the system becomes more dimerized, the SF to MI transition occurs
at smaller $U$ value. Note that, for nonzero $\delta$ value, both the SF and MI phases 
coexist with BW order, since nonzero $\delta$ inherently introduces bond energy fluctuations.
\begin{figure}
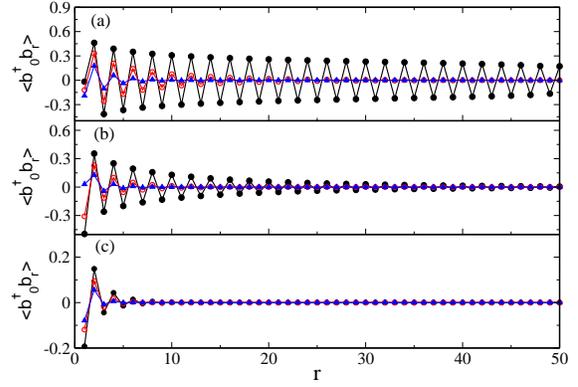

\onefigure[scale=0.3]{fig2.eps}
\caption{(color online) Plot of hopping correlation function against
$r=|i-j|$ for (a) $\delta=0.0$ (b) $\delta=0.5$ and (c) $\delta=0.8$. In each case the $U$
values are $0.5$ (filled circle), $3.0$ (open circle) and $5.0$ (triangle).}
\end{figure}

In Fig.~3 we present the asymptotic value of the order parameter, $<b_i^\dag b_j>^{1/2}$,
against $U$ for different $\delta$ values. In all cases the order parameter
vanishes (attains $\sim$ 10$^{-9}$)
at some critical $U$ values. The critical points have been determined through
opening up of the excitation gap, density fluctuations data and estimation of correlation length.
However, the error in estimating the critical points can be minimized if we 
determine the transition points where the asymptotic ($|i-j| \rightarrow \infty$) value of
$<b_i^\dag b_j>^{1/2}$ vanishes. In all cases, the
critical $U$ value deviates of the order of $\pm 0.1$. We find that with increase in $\delta$,
the order parameter vanishes at a smaller $U$-value. 
\begin{figure}
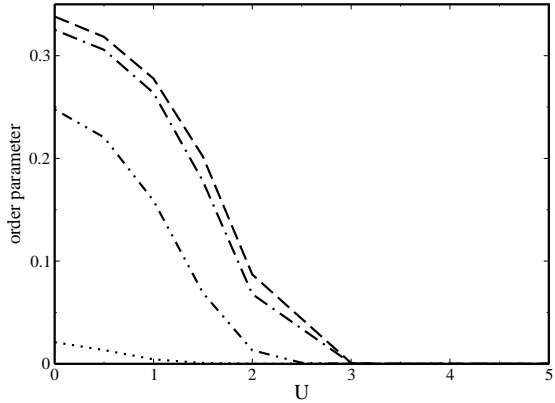

\onefigure[scale=0.3]{fig3.eps}
\caption{Plot of superfluid order parameter (asymptotic value of $<b_i^\dag b_j>^{1/2}$
such that $|i-j|\rightarrow \infty$) against $U$ for $\delta=0.0$ (dashed line),
$0.1$ (dashed dotted line), $0.3$ (dash dot dot) and $0.5$ (dotted line).}
\end{figure}

In Fig.~4, we present the $\delta$ vs. $U$ phase diagram. The critical line is shown with a width
which corresponds to the uncertainties in determining the critical $U$ values. 
The phase diagram shows that as the system becomes more dimerized the critical point
shifts to a lower $U$ value. 
Near the $\delta$-line beyond $\delta = 0.6$ the critical values of $U$ becomes very
small to determine. Infact, there exists an extremely narrow region beyond $\delta > 0.6$
where superfluid and bond-wave phase coexist and the phase boundary
touches asymptotically the point $\delta = 1$.
In fact, along $\delta$=1 line, the system essentially breaks into
disconnected dimers and the superfluid phase ceases to exist.
It should be mentioned clearly here that at $U=0$ and
$\delta=1$ (marked by a star
to avoid confusion with the filled circles which correspond to critical point for
transition) there is no phase transition, as along $\delta$=1 line the system consists
of dimers only.
The underlying physics is quite simple. The correlation
length or the characteristic length of the system decreases as we increase $\delta$ as the bond strength
alternates. Beyond $\delta = 0.6$, the correlation length becomes too small
to even estimate. If 
we consider a dimer (correlation length is smaller than $1$), it
corresponds
to $\delta=1.0$ in the Hamiltonian. For unit filling, the matrix eigen values are 
$U$, $\frac{U-\sqrt{32+U^2}}{2}$ and $\frac{U+\sqrt{32+U^2}}{2}$. Expanding the lowest excitation
gap for small $U$ gives an estimate of  $\frac{U}{2}+2\sqrt{2}+ \frac{\sqrt{2}U^2}{32}$ as the
gap value. Thus even for $U=0.0$, there is a finite gap
(for a dimer) which prevents SF phase to exist. In fact,
beyond $\delta = 0.6$, the correlation length becomes so small that
even a small (but finite) $U$ can drive the system to an insulating
phase.
In the phase diagram, the $\delta=0$ line (broken line) corresponds to pure SF phase below the 
critical point and above $U_c$, it corresponds to pure MI phase. However, for non-zero 
value of $\delta$, the phase transition is from superfluid-bond-wave (SF-BW)
phase to Mott insulator-bond-wave (MI-BW) phase. The main point
is that dimerization reduces the correlation length which in turn destabilizes the SF phase. Note
that, in this sense both the many-body repulsion and dimerization in kinetic energy drive the
system into insulating phase.
\begin{figure}
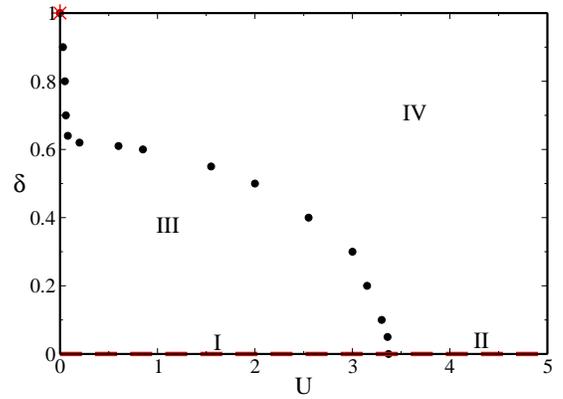

\onefigure[scale=0.3]{fig4.eps}
\caption{(color online) Phase diagram of $\delta$ vs. $U$. The $\delta=0$
line (broken line)
corresponds to SF to MI transition only. The different phases are (I) pure SF
phase, (II) pure MI phase (along the $\delta=0$ line), (III) SF-BW phase and
(IV) MI-BW phase. }
\end{figure}
\section{Conclusion}
In conclusion, we have studied the $\delta$-$U$ phase diagram of a system of bosons
in a one-dimensional optical lattice by density matrix renormalization group method.
Our investigation reveals the existence of a new phase, where superfluid phase
coexists with bond-wave phase due to the effect of dimerization
in hopping strengths. We have observed two types of phase transitions:
conventional superfluid to Mott insulator phase
and superfluid-bond-wave to Mott insulator-bond
wave phase. We have also calculated the critical
point at which the transitions occur. However, note that in our model
the bond-wave symmetry breaking is imposed ``externally" by the Hamiltonian itself.
Whether any new phase will appear due to the
effect of next nearest neighbor hopping is currently under study.

\acknowledgments
We acknowledge the research support from DST, India. SKP acknowledges
CSIR, Govt. of India and AOARD, US Airforce for research grant. 

\end{document}